________________________________________________________________________
\documentstyle[12pt]{article}
\begin{document}

\title {\bf{\Large Defect Formation and Crossover Behavior in the\\
Dynamic Scaling Properties of Molecular Beam Epitaxy}}
\author{S. Das\,Sarma, C. J. Lanczycki, S. V. Ghaisas and J. M.
Kim\\Department of Physics\\University of Maryland\\College Park,
Maryland  20742}
\maketitle
\begin{center}
{\large Abstract}
\end{center}
{\indent}Stochastic simulation results, appropriate for Molecular
Beam Epitaxy, involving ballistic deposition and thermally activated
Arrhenius diffusion of adatoms are presented for one- and
two-dimensional substrates, allowing for overhangs and bulk vacancies.
The asymptotic Kardar-Parisi-Zhang universality is
found to be triggered by a sudden nucleation of large-scale
defect formation in the growing film that shows a distinct dependence
on dimensionality.  The pre-nucleation
transient behavior, which may be of experimental relevance due to the
low defect content, is associated with standard solid-on-solid
universality classes.

\vfill
\newpage

There has been considerable recent
theoretical
\cite{dst,lds,wv,vil,gb,yan,kls,spl,sdg,kot,wilb,dst2,gds2,kps,alf,dsgk,psw,vst} and
experimental \cite{eag,chev,ylw,he,cotta}
activity in understanding Molecular Beam Epitaxy (MBE) as a
statistically scale invariant phenomenon involving the dynamical evolution
of a self-affine fractal interface \cite{dyn}, which roughens kinetically
during
growth due to the random noise associated with the deposition of an
incident particle flux.  In particular, growth exponents \cite{dyn},
characterizing the self-affine roughness of the growing MBE interface,
have been extensively studied using coarse-grained continuum
theoretical models and renormalization group techniques, as well as
discrete atomistic growth models and numerical simulations.  Recently,
a number of experimental results \cite{eag,chev,ylw,he,cotta} have reported
measurements of the MBE growth
exponents, claiming agreement with one or the other of the
theoretical models.  It is safe to say that no consensus has emerged
on the dynamic universality class of MBE growth, either among
theoretical papers which often employ different models, or, among
experimental results which use different materials, growth
conditions and measurement techniques. \par
The purpose of this article is to report numerical results of a
MBE simulation which attains more realism by going beyond
the existing calculations in the literature in eliminating all the
non-essential approximations:  the solid-on-solid
(SOS) approximation, violation of detailed balance, infinite vertical
diffusion, and instantaneous relaxation.  Our model instead
allows overhangs and bulk vacancies,
and treats the diffusion process dynamically as a
thermally activated Arrhenius
hopping process between neighboring sites involving the breaking of
chemical bonds.  The only essential approximations of our model are
the assumption of a discrete lattice gas system ({\it i.e.} the atoms
are allowed only at pre-defined lattice sites which are taken to form
a $d$-dimensional hypercube), and the simulation of atomistic kinetics with a
stochastic Monte Carlo algorithm where the probability of the
occurrence of a particular kinetic process is determined by an
Arrhenius thermally activated process.  Growth will always occur along
the vertical direction on a $d'=d-1$ dimensional substrate.
A non-essential approximation of
our model is the neglect of evaporation.\par

The novel results reported herein are the following.
We find that the
transition to the asymptotic Kardar-Parisi-Zhang (KPZ) \cite{kpz} universality
(expected \cite{dyn} because we allow formation of overhangs and vacancies in
our models)
occurs via an {\it unexpectedly} rapid formation of a fairly high
level of bulk defect density in the growing film for all models studied.
Furthermore, the defect density manifests qualitatively
different behavior in $d'=1$ and $d'=2$.
At early times, before this nucleation takes place, few defects are
present in the bulk and a crossover region emerges which may correspond
to various non-KPZ dynamic universality classes
appropriate to defect free models without
evaporation despite the eventual formation of
defects.  The predicted KPZ universality manifests itself only after
the density of defects has saturated.
We attribute the disagreement \cite{eag,chev,ylw,he,cotta}
among various experimentally measured MBE growth
exponents to the possibility that different measurements may be
exploring different crossover regimes during MBE growth.\par

We analyze the simulated MBE dynamical
interface width, $W(L,t)$
({\it i.e.} the root mean square fluctuation in the surface
height), as a function of the growth time $t$ for fixed
substrates of size $L^{d'}$ for $d'=1$ and $2$.  The standard
hypothesis \cite{dyn} is that, $W$
exhibits dynamical scaling behavior:

\begin{equation}
W(L,t) \sim L^{\alpha}f(\frac{t}{L^{z}}),
\end{equation}

\noindent where $f(x)\sim x^{\alpha/z}$ for $x \ll 1$ and $f(x)\sim$ constant
when
$x \gg 1$, leading to

\begin{equation}
W(L,t) \sim \left\{ \begin{array}{ll}t^{\beta}&\mbox{for $t^{1/z}\ll L$}\\
                                  L^{\alpha}&\mbox{for $t^{1/z}\gg L$}\\
\end{array} \right
\end{equation}

\noindent with $\beta=\alpha/z$, and $\alpha$, $\beta$ and z are respectively
called the roughness, growth and dynamical exponents.
 From our simulation of
$W(L,t)$ we comment on the {\it effective} critical exponent $\beta$ of
MBE growth as a function of both the growth time and the
temperature, $T$, which, by
controlling the diffusion rate through the activated hopping of the
atoms, becomes the dominant determinant of the crossover behavior.  In
all our results, $W$ is measured in number of layers ({\it i.e.}
lattice spacing) and $t$ is measured as the number of defect-free
layers worth of
material deposited given that $R_{dep}$, the deposition rate, remains
fixed throughout at 1 layer/sec.\par

We employ a ballistic deposition \cite{dyn} of atoms
onto the surface, abbreviated {\bf BD}.  Here, an atom drops
vertically above a randomly chosen
surface site, being added to the crystal at the first encounter with
an occupied nearest neighbor (NN) site, in contrast with SOS models
where deposited atoms always land atop the particle directly below
the falling atom.
In general, hopping will also lead to vacancies in the aggregate and
this type of hopping is called ``ballistic,'' notated as
{\bf BH}: both surface and bulk atoms experience this thermal motion,
hence all defects throughout the crystal have the potential to heal.
Most of our models combine {\bf BD} and {\bf BH} as given above,
developing defects both in
the deposition process and via hopping, and will be called the {\bf
BD/BH} model.  Note that the term ``defect'' means only a ``vacancy''
({\it i.e.} the absence of an atom at a lattice site) within our
lattice gas model -- we can have both bulk and surface vacancies in
the {\bf BD/BH} model.  The SOS model, on the other hand, does not
allow any defects.\par

A continuum equation such as the KPZ
equation \cite{kpz} should govern the asymptotic characteristics of
the growth in the {\bf BD/BH} model due to the presence of defects
\cite{dyn}, introducing a current
non-conserving nonlinear term, $\lambda(\nabla h)^{2}$, inexpressible
as the divergence of a surface current due to
the sideways growth possibilities in {\bf BD}.  The height,
$h(\vec{x},t)$, is defined as the
distance to the highest occupied lattice site directly
above the substrate coordinate $\vec{x}$.
For one substrate
dimension, the KPZ equation predicts that $\beta = 1/3$, while
$\beta \approx 1/4$ when measured for $d'=2$
\cite{tang}. One might expect that at
higher temperatures, when the density of the crystal approaches unity,
current non-conserving effects disappear and a conservative
description of growth extensively studied in SOS models
\cite{dst,lds,wv,vil,sdg,wilb,dst2,ew}, could become valid.  We define
``conservative'' as the $\lambda=0$ case of a general continuum
equation proposed for MBE growth \cite{dst2,vil}, for all the remaining
$h$-dependent terms are divergences:

\begin{equation}
\frac{\partial h}{\partial t} = \nu_{2}\nabla^{2}h -
\nu_{4}\nabla^{2}(\nabla^{2}h) + \lambda(\nabla h)^{2} +
\lambda_{2}\nabla^{2}(\nabla h)^{2} + \lambda_{3}\nabla(\nabla h)^{3}
+ \eta
\end{equation}

\noindent where $\eta$ is an uncorrelated, Gaussian noise term.  When
$\nu_{2}\ne 0$ and $\lambda = 0$, the width scales as described by
Edwards and Wilkinson (EW) \cite{ew}, while the fourth order terms
$\nabla^{2}(\nabla^{2} h)$ (initially considered by Das\,Sarma and
Tamborenea and Wolf and Villain \cite{dst,wv})
and $\nabla^{2}(\nabla h)^{2}$ (introduced by Lai and Das\,Sarma and
Villain (LDV) \cite{lds,vil}) become relevant terms in the
$\lambda = 0$ limit when $\nu_{2}=0$.  We will not have much to say about the
$\lambda_{3} \nabla(\nabla h)^{3}$ term, which is a higher order
correction to the linear EW second order term and was first introduced
by Lai and Das\,Sarma \cite{lds}.\par

Central to the implementation of the {\bf BD/BH} model
is a bond dependent activated
hopping \cite{crc,mar,kkd}, with $E_{n} = 1\, eV + (.3\, eV)n$, where $n=1,2$
or 3 is the number of
occupied nearest neighbor sites
(particles with four nearest neighbors
are considered to be unhoppable).  This defines a set of modified
Arrhenius hopping rates of the form $R_{n}=\frac{d'k_{B}T}
{{\rm h}}e^{-E_{n}/k_{B}T}$, where the constants in $E_{n}$ were
semiempirically chosen to correspond to the energetics of silicon and
${\rm h}$ is Planck's constant.
Note that with
Arrhenius activated hopping, detailed balance is explicitly
obeyed because the activation energy depends only on the initial
state.\par

We study a wide range of temperatures, with the ratio
$r = R_{1}/R_{dep}$ varying from 0.03 at $T=450 K$ to $\sim 6300$ at
$T=700 K$, and $R_{2} \ge R_{dep}$ at $T \ge 610 K$.  The parameter
$r$, which depends only on the temperature,
is crucial in the simulation, measuring the relative strength of
the two dynamical processes in our simulation:  deposition and
diffusion.  For $r \ll 1$, deposition is the fastest process and
the diffusion length is $\ll L$.  On the other hand, when singly
bonded atoms hop faster than the rate at which
atoms are added to the surface, $r \gg
1$.  While the atom has a larger diffusion length and can explore more
of the surface before a layer is deposited, $r$ only measures hopping
from singly bonded sites:  when the particle reaches a doubly bonded
site, the rate $R_{2}$ becomes relevant, with an exponentially
suppressed diffusion length.  Thus $r>1$
leads to the rapid traversal of terraces, but
unless $R_{2} > R_{dep}$, inter-terrace hopping is
slow and the diffusion length is of the order of the
average terrace size.  Of course, as $r$ ({\it i.e.} T)
increases, the surface atoms more easily diffuse and this model should
smooth out more and more of the defects.  At extremely high
temperatures, where $r\gg1$, defect formation becomes difficult, and we
expect this
model to generate an approximately solid-on-solid crystal.  By increasing
$r$, therefore, we hope to observe conservative growth (when there are
very few defects) as a crossover
prior to the non-conserved universality
class (which should always be the asymptotic class as $\lambda$ is a
relevant parameter in a renormalization group sense.)\par

Unless specified, atoms
are allowed to hop to any nearest neighbor (NN) or next-nearest
neighbor (NNN) site with equal probability provided that the landing
site provides at least one NN bond:  we do not impose any restrictions
on the change in coordination due to a hop.  In $d'=1$, there are 8 potential
landing sites while in $d'=2$ a hopping atom may move in one of 18
directions, the topology providing for more robust diffusion.
While this aspect of
our model ($d'=1$) is similar to that of Kessler et. al. \cite{kls},
our implementation (described below)
removes the use of unphysical instantaneous relaxation, with the accompanying
detailed balance violations, for closer links with experimental work.
Our $d'=2$ results are totally new.\par

Because bulk voids are permitted in this study, we must keep
track of not only the surface particles (sufficient in SOS
simulations) but large portions of the bulk, leading to formidable memory
requirements, especially in $d'=2$.  This required us to periodically
discard the lower portions of the bulk crystal whose activity was confirmed
to no longer influence the surface properties.
Additionally, as early times ($< 100$ layers) in the growth
process were analyzed, we
utilized a Monte Carlo algorithm which
permitted diffusive moves more often than just upon the deposition of
a new particle whenever the hopping of singly bonded atoms was the
fastest process ({\it i.e.} when $R_{1}>R_{dep}$),
to avoid graininess in our data.
For $T > 500$\,K where $r > 1$, the internal time increment
changed \cite{kkd} from the time to deposit a single
particle to being the shortest hopping
period, $\tau_{1}=\frac{1}{R_{1}}$.  Thus,
at $T=700$\,K, for example, where $R_{1} \approx 6300\,R_{dep}$, we allowed
$6300$ opportunities for singly bonded particles to
hop between deposition events.
Of course, the number of hops actually executed was much smaller and limited
by the number of appropriately bonded atoms available
at each instant.  Because searching large arrays
with such frequency proved highly prohibitive from a time standpoint, we stored
a lookup table providing the identity and
coordination of all particles eligible to hop,
updating the hopping status and coordination of any affected atoms
after each movement within the crystal.  A newly deposited
particle with a single bond, for example, continued to
diffuse (quickly) to neighboring singly bonded
sites until randomly selecting a more highly bonded landing
site, thus changing its coordination in the stored table and
rendering it ineligible to hop further as a singly bonded atom.  The
selection of a hopping particle from a number of eligible
particles was entirely random, and the thermal hopping rates are simply average
rates in that individual particles may hop more or less rapidly
depending on the local morphology of the surface.\par

Figures 1a and 1b show how the surface width varies in time for $d'=1$
and $2$, respectively, for the {\bf BD/BH} model with both upward and
downward hopping allowed.
After an initial transient of
$\beta\,=1/2$ due to uncorrelated growth (not shown), the surface
profiles develop regions of smaller slope which increase in duration and
flatten with temperature.  Intervening between the early slow growth
in the width and the asymptotic scaling behavior, which (for $d'=1$) Fig. 1a
shows to be reasonably close to the KPZ prediction, is
a quite dramatic increase in
the width where the effective value of $\beta$ clearly exceeds the late time
value.  The rapid development of surface height fluctuations
manifests itself most strongly in the physically relevant $d'=2$
case.  To determine
the origin of this observation and to discover any implications
such an occurrence might have for crystal growth studies is our main focus in
this paper.\par

At early growth times, the $d'=1$ case (Fig. 1a)
indicates the gradual emergence of a transient region as the
temperature rises, with a slope
$\beta\approx 0.25$ at $T=700$\,K
(comparable to that predicted for defect free aggregation via EW growth).
For $d'=2$, there
exists an even more pronounced linear region whose slope
decreases with temperature.  The $d'=2$ EW equation manifests a weak
scaling behavior, predicting a logarithmic divergence with time:  at
$T=680$\,K, $\beta < 0.1$ up to $t \sim 100$ layers, which is
consistent with the presence of logarithmic scaling.\par

At low $T$ no early time extended region of low ($< 1/2$)
slope is apparent, although in all cases the slope does
inflect slightly prior to a sharp rise of $W$.  In the range
($t \sim 1-100$ layers), the interface goes from being dominated by a sharp
accumulation of fluctuations when diffusion is slow ({\it i.e.} low
$T$) to yielding a plateau with a small rate of increase in the width
at high $T$ before a rapid jump in $W$.
Thus increased temperature can evidently
postpone the rapid growth of $W$ in time; ultimately, at high $T$,
giving rise to a plateau with an
effective $\beta$ quantitatively in line with the various
possible predictions of Eq. 3 for small growth times.
Interestingly, in $d'=1$ raising the temperature makes the accumulation of
fluctuations milder whereas it results in a more dramatic jump in the
width in the $d'=2$ simulation.  From Figs. 1a and 1b, the slope
of the transition region is $\sim .4-.6$ in $d'=1$ while $d'=2$ indicates an
effective slope $\geq 1$, showing this strong influence of
dimensionality.\par

To help discern the cause for this precursor to the asymptotic
universality,
we refer to Figs. 2a and 2b, where we display the average growth
velocity, $\overline{v}(t)=\partial \overline{h}/\partial t$,
in time in $d'=1$ and $2$, respectively.
If there exist no voids, $\overline{v}\,\equiv\,1$ layer/sec,
and all deviations from
unity directly reflect the accumulation of defects in the growing
film through the relation $\overline{v} = 1/(1 - DD)$, where $DD$
represents the bulk defect density ({\it i.e.} the
fraction of the bulk crystal which is empty).  Notice the
sharpness with which velocity changes around
$t\sim 100-300$ in $d'=2$, showing
a constant value, $\overline{v}_{\rm sat}$, at
long times.  The behavior in $d'=1$ is
qualitatively similar, but not as sharp.
Further, the onset and temporal range of rapid changes in $\overline{v}$
closely track the strong variation in
width for the corresponding plots in Fig. 1.  Finally, the dashed lines
in Fig. 2 show that $DD$, the bulk
defect density (a bulk defect being defined as any vacant site
in the inactive region of the crystal),
also follows the general behavior of $W$, clear proof that the
sharp rise in width is intimately related to a very
sudden onset of bulk defect formation.
Results shown in Figs. 1 and 2 establish that the asymptotic
scaling regime is necessarily triggered by the sudden (not gradual)
formation of
a very large bulk defect density, and the onset of large scale defect
formation shifts to later times as
the temperature increases ({\it i.e.} the thickness of the defect-free
part of the growing film increases with temperature).
Our $d'=1$ simulation at low temperature
was the only case that allowed for a good measurement of the
asymptotic $\beta$, producing an asymptotic slope near $0.29 \pm .01$,
in acceptable agreement with other (less complicated) simulations
allowing overhangs \cite{dyn}.  Despite the inability to perform
long (and large) enough simulations to measure the asymptotic
$\beta$ explicitly
for $d'=2$ (or at high $T$ in $d'=1$), the existence of a non-zero
defect density implies that the coefficient $\lambda$
in Eq. 3 is finite, and therefore in the asymptotic limit the KPZ
universality class must always control growth.  Thus, we
take the asymptotic scaling to be KPZ based on the presence of
defects in all cases.  Computational limitations
precluded more quantitatively accurate measurements of the slope
$\beta$ for early times, because we could not increase $T$ sufficiently
to broaden those regions further (this was a greater problem in
$d'=1$, however).\par

The appearance of defects and KPZ scaling only after strong
defect nucleation leads to an interesting possibility at high
temperatures, namely that
there exists a regime during ballistic deposition in which
the density of defects is near zero and insufficient to yield the
non-conservative growth typified by the KPZ equation.  We therefore
interpret this period in the growth as a crossover regime that
obeys a conservative continuum equation ({\it i.e.} a $\lambda=0$
version of Eq. 3) which usually is the exclusive domain of
solid-on-solid simulations.  In a single run, unfortunately,
we were not able to see both the early defect-free regime and
the asymptotic, KPZ scaling due to computational limitations, as
mentioned above.  At our highest temperature in $d'=1$ ($T=
700$\,K), the slope in the early defect-free region is $\sim 0.245 \pm 0.005$,
consistent with the EW ($\nu_{2} \ne 0$) prediction of $\beta = 1/4$
in $d'=1$.
Physically, the diffusion to local height minima can be described
by the EW equation, and this is the dominant mode of diffusion
at early times where the width is $\le 1$ and the height fluctuations
are too small to support voids:  hence it seems reasonable
to observe EW behavior to manifest at early times.
Should SOS-like, low defect growth persist for long times
at yet higher temperatures, it has been
seen \cite{wilb} that LDV behavior ($\lambda_{3}=\lambda=\nu_{2}=0$,
$\lambda_{2} \ne 0$) could emerge:  some of the experiments
\cite{he,cotta} would support this possibility.\par

This result generally allows a low-defect crystal to grow mimicing
solid-on-solid aggregation for a finite time even though the underlying
growth rules are more general and allow voids in the bulk.  In
fact, an initial dip in the instantaneous slope of the width versus time
plots has occurred in disparate cases \cite{dst,kls,wilb,dst2} in the
literature, all of which employ some sort of diffusion length
in addition to the length scale set by the system size.
An early, transient scaling region thus may be a general phenomena
in growth with competing dynamical processes.\par

Recent experimental work of Eaglesham \cite{eag} and our
studies in $d'=2$ here both suggest a nucleation of defects in MBE
growth, which seems to happen once enough configurations involving
voids and overhangs are available on the surface.  This
would predict an exponential dependence
of the turn-on time of the defect formation, $t_{\rm c}\sim$ exp[$-E_{\rm
a}/k_{\rm b}T$], where $E_{\rm a}$ is an effective activation energy
for creating a defect.
In $d'=1$, we find that this hypothesis indeed holds, with a defect
activation energy of $1.1 \pm 0.2\,eV$.  The same $E_{\rm a}$ was
found from a similar fit in $d'=2$.  Eaglesham inferred $E_{\rm a}$
to be $.4 - 1.5\,eV$, dependent upon the growth
rate on Si(100).  We believe that the sudden onset of defect formation (with
crossover to KPZ) found in our simulations is related to Eaglesham's
observations \cite{eag} of a limiting epitaxial film thickness and
$\beta \approx 1$.  We emphasize that our results clearly show that
the thickness of the defect-free film depends strongly (exponentially)
on temperature.\par

Looking at the saturation defect density as a function of temperature
we find a surprising dichotomy between the $d'=1$ and $2$ models.  One
would naively expect that as hopping becomes more effective at higher
temperatures in
eliminating defects, the defect density (DD) should drop to
zero eventually.  In fact, this prediction is borne out for $d'=1$:  at
$T=0\,K$, $DD=1/2$ ($\overline{v} \rightarrow 2$) while at
$T=700\,K$, $DD\approx .01$ ($\overline{v} \rightarrow 1$).
In $d'=2$, $DD(T=0\,K) \approx 2/3$, but
referring back to Fig. 2b, $DD(T=620\,K) \approx 2/3$
as well.  In both cases, we
have found nucleation type behavior, but for $d'=1$, $\overline{v}$ rises
sharply from unity to a decreasing, $T$-dependent constant while in $d'=2$
the defect density seems to always attain the large $T=0$
value of $2/3$, doing so more rapidly than the $d'=1$ model.  Defect
formation is thus substantially easier in $d'=2$, for topological
reasons which we feel are intimately linked with the extra degrees of
freedom associated with atomic diffusion on a 3-dimensional
lattice, as compared with diffusion in $d=2$ dimensions.  Despite this,
however, one expects growth in both dimensions to display asymptotic scaling
behavior appropriate for the KPZ equation,
transcending this morphological difference.  As emphasized above, this
asymptotia sets in only after the growth of a defect-free film whose
thickness is determined by temperature and whose growth dynamics seems
to be describable in terms of conserved SOS growth dynamics.\par

Because the atoms in
our simulated crystals are forced by the algorithm to always reside at
lattice sites, large defect densities $\ge 1/2$ can arise.  In such an
open structure, defining the ``surface'' as the highest point of each
column may be a problem as there are numerous active perimeter sites
that are at lower heights.   Of course, in a
real crystal a lattice energetically cannot support the increased
surface area that a large
density of voids requires, and the growing crystal would become
amorphous at some point.
In this sense, then, the dimensionality dependence of $DD$ is an academic
point, but the rapid accumulation of defects may have a bearing on the
suddenness of a crystaline-amorphous transition, the strength of which
may retain the dimensionality dependence.  Generally, this entire
question of how to define a surface for a crystal with a high defect
density and describe
it analytically remains an unresolved issue of scientific interest.
Our approach in this paper has been the standard one \cite{dyn}
assuming that a single-valued height, $h$, can be uniquely defined.\par

To investigate the prevalence of the nucleation threshold for defect
formation, we have also carried out a series of simulations suppressing any
upward hopping in our {\bf BD/BH} MBE growth models.
We summarize our findings for the
``no up hopping'' model here:
\begin{enumerate}
\item In $d'=1$, the no-up-hop {\bf BD/BH} model substantially reduces
the sharp rise of the defect formation, prolonging instead the initial
flat (EW) scaling regime, and producing prolonged layer-by-layer growth
oscillations at the highest ($\sim700\,K$) temperatures.  The
asymptotic long time value of $\beta$ remains $\approx 0.3$, being consistent
with the KPZ value of $1/3$, but it gets there gradually in time instead of
going through a rapid defect formation phase.
\item In $d'=2$, the no-up-hop rule
more clearly points towards EW type ($w \sim ln(t)$) initial scaling,
followed again by strong defect nucleation, with the saturated
defect density still being $\approx 2/3$.  The
ballistic nature of our model is too robust in $d'=2$
to be affected by the
no-up-hopping rule.  We reiterate our belief that there is a
substantial difference between $d'=1$ and $2$ in terms of the
available defect and hopping configurations, and the different
topologies in the two cases make it dangerous to draw detailed
quantitative conclusions
about real ({\it i.e.} $d'=2$) MBE growth, where voids are rare (but
permitted), based on one-dimensional
simulations.
\end{enumerate}
\par
Most of the large scale MBE simulations \cite{dst,wilb,dst2,vst,crc,kkd}
explicitly make the SOS approximation, with the exception of one recent
simulation \cite{mar}, whereas it might be emerging as the high temperature,
early time limit of our fully ballistic model.   We
note that experiments ($d'=2$) have observed $\beta$ from low values $\approx
0.2-0.3$ \cite{chev,he,cotta} to up to the unstable case of $\beta \sim 1$
\cite{eag}, and our model provides the following
scenario which could unify these
observations.  After the sub-monolayer Poisson region of uncorrelated
random growth, $t_{c}$ layers of
nearly defect-free material can be deposited, where $t_{c}$ is an
epitaxial thickness \cite{eag,dst2} having an
exponential dependence on $T$.  Based on $d'=1$ measurements, the
slope of this early time, pre-nucleation growth appears
to approach the EW value for $\beta$ at our highest temperature.
Of more relevance, in $d'=2$ we
believe that the EW universality also is the extreme non-equilibrium
limit, prior to defect formation, with other conservative
universality classes possibly manifesting transiently at temperatures
where diffusion to height minima is not the dominant process.

After $t_{c}$ layers, the surface can support enough centers about which to
sustain a nucleation of defects, which occurs at a
high rate ($\beta_{eff} \sim 1.0$ in $d'=2$) and with an
intriguing $d'$-dependence.
The surface incorporates as many defects per layer as it
can sustain and the defect density saturates fairly fast.
Once this time independent
defect density emerges, the asymptotic KPZ scaling nature of the surface
develops as the KPZ nonlinearity parameter, $\lambda$, which is
related to $\overline{v}_{\rm sat}$, attains a steady state
value:  the constancy of $\lambda$ is an important
prerequisite for claiming KPZ universality to be present.
Measurements made within either the KPZ, or early time,
high-$T$ pre-nucleation regimes will show small $\beta$ ({\it i.e.}
$\beta <.5$), while unstable ($\beta > .5$)
growth could be inferred by observing
the nucleation region itself.
The sort of complexity which results here from the interaction among
processes with competing time scales makes any
clearcut observations of specific dynamical universality classes
extremely challenging experimentally (and simulationally).\par

We should note that while, in general, values
of $\beta > 1/2$ typically reflect an
instability of the growth model, such nomenclature does not fit this
model, for after $\overline{v}$ saturates, $\beta$ attains values less
than $1/2$ (on the order of $0.29$ in $d'=1$).  A truly unstable
growth model would manifest no such turnover:  the width would grow
continuously without bound with $\beta > 1/2$.  Even in those
models for which the
asymptotic scaling cannot be measured, there is a turnover in $W$
which signifies the model's ultimate stability.\par

To the best of our knowledge, there has not been any earlier attempt
at the full {\bf BD/BH} MBE growth simulations (either in $d'=1$ or
$2$) using temperature dependent Arrhenius activated hopping.
We believe, though, that {\bf BD} may actually over-emphasize the proclivity of
defect formation during deposition (as seen by the defect nucleation even
when up hopping was suppressed), and real MBE
deposition may be intermediate between {\bf BD/BH} and the more usual SOS
deposition rule, depending on the local solid state chemistry.  In
preliminary runs where we adjust the sticking probability down to zero
for depositing particles (while it remains unity for atoms which hop)
defect formation no longer occurs suddenly and the pre-KPZ region
at lower temperatures may be more amenable to a quantitative analysis
in terms of the universality classes of Eq. 3.\par

To summarize, the major result of this study is the identification of
a rapid, rather than gradual, onset of defect formation in a
realistic ballistic deposition Arrhenius model
for crystal growth, displaying both conservative
and non-conservative characteristics at different times during
the crystal's growth.  A qualitative difference in the saturation
defect density behavior with dimensionality was discovered:
in $d'=2$, the defect density approached $2/3$ for
all $T$ while on a one-dimensional substrate, it decreased
monotonically to zero.  The conservative universality class that seems to be
associated with the early time region at high $T$ is that of EW,
with the expected
KPZ universality taking over after defect nucleation.
These simulations could
provide a consistent explanation for many differing experimental
results which have so far defied explanation.\par
This work is supported by the US-ONR.\par

\pagebreak

\begin{center}
{\bf{\large FIGURE CAPTIONS}}
\end{center}
\noindent Fig. 1:  (a) Shows log-log plots for $W(L,t)$ vs. $t$ in
$d'=1$ dimension for the temperatures $450$ (------), $550$ ($\cdots$
$\cdots$), $600$ (-- -- --) and $700\,K$ ($\cdot$-$\cdot$-$\cdot$).
The system size is 2000 lattice sites, except $T=700\,K$ where it is
10000. The guide lines have slopes of 1/4 and $\sim$ 0.29. (b) Shows
$d'=2$ dimension results with $500$\,x\,$500$ substrate at
the temperatures $580$ (------), $620$ (--
-- --), $640$ ($\cdot$-$\cdot$-$\cdot$) and $680$\,K
(--$\cdot \cdot \cdot$--).\par

\vspace*{1cm}

\noindent Fig. 2:  (a) Plot of the growth velocity $d\bar{h}/dt$
(------) and bulk defect density (-- -- --) measured directly from the
simulation in $d'=1$.  $T=700$\,K and $L=10000$
for a single run.   (b) Same plot in
$d'=2$ for a $500$\,x\,$500$ system at $T=620\,K$, but the bulk defect
density was calculated from $(1 -(t/\bar{h}))$.\par

\end{document}